\documentclass{article}[13pt]
 \usepackage{latexsym}
 \usepackage{amsmath}
 \usepackage{amsfonts}

\usepackage[T1]{fontenc}
\usepackage[utf8]{inputenc}

\usepackage[unicode=true]{hyperref}
 \hypersetup{
   pdfstartview={FitH},
   pdffitwindow=false,
   pdfborder={0 0 0},
   colorlinks=true,
   linktoc=page,
   linkcolor=blue,
   urlcolor=blue,
   hypertexnames=false,
   pdfdisplaydoctitle=true,
   pdfduplex=DuplexFlipLongEdge,
 }

\input{amssym.def}
\input{amssym.tex}

\newcommand\dist{\mathrm{dist}}

\newtheorem{theoreme}{Theorem }[section]
\newtheorem{proposition}[theoreme]{Proposition}
\newtheorem{lemma}[theoreme]{Lemma}
\newtheorem{definition}[theoreme]{Definition}

\newtheorem{remark}[theoreme]{Remark}

\newtheorem{open}[theoreme]{Open Problem}
\newcommand{\beq}{\begin{equation}}
\newcommand{\eeq}{\end{equation}}

\def\bel{\begin{lemma}}
\def\eel{\end{lemma}}
\def\bet{\begin{theoreme}}
\def\eet{\end{theoreme}}
\def\bed{\begin{definition}}
\def\eed{\end{definition}}
\def\bep{\begin{proposition}}
\def\eep{\end{proposition}}
\def\ber{\begin{remark}}
\def\eer{\end{remark}}
\newcommand{\R}{\mathrm{R}}
\newcommand{\I}{\mathrm{I}}

\newcounter{smallarabics}
\newenvironment{arabicenumerate}
{\begin{list}{{\normalfont\textrm{(\arabic{smallarabics})}}}
  {\usecounter{smallarabics}\setlength{\itemindent}{0cm}
   \setlength{\leftmargin}{5ex}\setlength{\labelwidth}{4ex}
   \setlength{\topsep}{0.75\parsep}\setlength{\partopsep}{0ex}
   \setlength{\itemsep}{0ex}}}
{\end{list}}

\newcounter{smallroman}

\newcommand{\ben}{\begin{arabicenumerate}}
\newcommand{\een}{\end{arabicenumerate}}

\def\rr{{\mathbb R}}

\def\cc{{\mathbb C}}

\def\bbbone{{\mathchoice {\rm 1\mskip-4mu l} {\rm 1\mskip-4mu l}
{\rm 1\mskip-4.5mu l} {\rm 1\mskip-5mu l}}}
\def\one{\bbbone}

\def\oplusal{\mathop{\hbox{\raise 1.5 ex
  \hbox{$\scriptscriptstyle\rm al$}
\kern -0.92 em \hbox{$\oplus$}}}}
\def\otimesal{\mathop{\hbox{\raise 1.5 ex
  \hbox{$\scriptscriptstyle\rm al$}
\kern -0.92 em \hbox{$\otimes$}}}}
\def\Gammal{\hbox{\raise 1.68 ex 
\hbox{$\scriptscriptstyle\rm al$}\kern -0.50 em $\Gamma$}}

\def\bar{\overline}

\def\c{{\rm c}}

\renewcommand\Re{{\rm Re}}

\def\i{{\rm i}}

\def\supp{{\rm supp}}

\def\e{{\rm e}}
\def\d{{\rm d}}
\def\Ran{{\rm Ran}}

\def\cD{{\cal D}}
\def\cG{{\cal G}}
\def\cH{{\cal  H}}

\def\cV{{\cal V}}
\def\cF{{\cal F}}

\def\cV{{\cal V}}

\def\cQ{{\cal Q}}
\def\cK{{\cal K}}

\def\12{\tfrac{1}{2}}
\def\32{\tfrac{3}{2}}
\def\52{\tfrac{5}{2}}

\def\qed{$\Box$\medskip}
\def\proof{\noindent{\em Proof.}\ \ }

\usepackage[normalem]{ulem} 
\def\ch{\mathcal{H}}
\usepackage[usenames,dvipsnames,svgnames,table]{xcolor}

\begin{document}
\title{On the domains of  Bessel operators}

\author{Jan Derezi\'{n}ski\\Department of Mathematical Methods in
  Physics,\\
  Faculty of Physics, University of Warsaw, \\Pasteura 5,\\ 02-093
  Warszawa, Poland,\\email: jan.derezinski@fuw.edu.pl\\ \\
  Vladimir Georgescu\\Laboratoire AGM, UMR 8088 CNRS,\\
  CY Cergy Paris Universit\'e,\\ F-95000 Cergy, France, \\
  email: vladimir.georgescu@math.cnrs.fr}
\maketitle

\begin{abstract}
  We consider the Schr\"odinger operator on the halfline with the
  potential $(m^2-\frac14)\frac1{x^2}$, often called the Bessel
  operator. We assume that $m$ is complex.  We study the domains of
  various closed homogeneous realizations of the Bessel operator.  In
  particular, we prove that the domain of its minimal realization for
  $|\Re(m)|<1$ and of its unique closed realization for $\Re(m)>1$
  coincide with the minimal second order Sobolev space. On the other hand,
    if $\Re(m)=1$ the minimal second order Sobolev space is a subspace
    of infinite codimension of the domain of the unique closed
     Bessel operator. The properties of Bessel operators are compared with
  the properties of the corresponding bilinear forms.
\end{abstract}



\section{Introduction}


\subsection{Overview of closed realizations of the Bessel operator}

The Schr\"odinger operator on the half-line given by the expression
\begin{equation}
  L_\alpha := -\frac{\d^2}{\d x^2} + \Big{(} \alpha - \frac{1}{4}
  \Big{)} \frac{1}{x^2} \label{eq-L-m-}
\end{equation}
is often called {\em the Bessel operator}.  The name is justified by
the fact that its eigenfunctions and many other related objects can be
expressed in terms of Bessel-type functions.

There exists a large literature devoted to self-adjoint realizations
of \eqref{eq-L-m-} for real $\alpha$.  The theory of closed
realizations of \eqref{eq-L-m-} for complex $\alpha$ is also
interesting.  Let us recall the basic elements of this theory,
following \cite{Derezinski1,Derezinski2}.

For any complex $\alpha$ there exist two most obvious realizations of
$L_\alpha$: the {\em minimal} $L_\alpha^{\min}$, and the {\em maximal}
$L_\alpha^{\max}$.  The complex plane is divided into two regions by
the parabola defined by
\begin{equation}\label{para}
  \alpha=(1+\i\omega)^2, \quad \omega\in\rr,
\end{equation}
(or, if we write $\alpha=\alpha_\R+\i\alpha_\I$, by
$\alpha_\R+\sqrt{\alpha_\R^2+\alpha_\I^2}=2$). To the right of this
parabola, that is, for $ |\Re\sqrt\alpha|\geq1$, we have
$ L_\alpha^{\min}= L_\alpha^{\max}$.  For $|\Re\sqrt\alpha|<1$, that
is to the left of \eqref{para}, $\cD(L_\alpha^{\min})$ has codimension
$2$ inside $ \cD(L_\alpha^{\max})$.  The operators
$ \cD(L_\alpha^{\min})$ and $ \cD(L_\alpha^{\max})$ are homogeneous of
degree $-2$.

Let us note that in the region $ |\Re\sqrt\alpha|<1$ the operators
$L_\alpha^{\min}$ and $ L_\alpha^{\max}$ are not the most important
realizations of $L_\alpha$.  Much more useful are closed realizations
of $L_\alpha$ situated between $ L_\alpha^{\min}$ and
$ L_\alpha^{\max}$, defined by boundary conditions near zero. (Among
these realizations, the best known are self-adjoint ones corresponding
to real $\alpha$ and real boundary conditions). All of this is
described in \cite{Derezinski1}.

Among these realizations for $\alpha\neq0$
only two, and
for $\alpha=0$ only one, are homogeneous of degree $- 2$. All of
them are covered by the holomorphic family of closed operators $H_m$,
introduced in \cite{Derezinski2} and defined for $\Re(m)>-1$ as the
restriction of $L_{m^2}^{\max}$ to functions that behave as
$x^{\frac12+m}$ near zero. Note that
\begin{align}\label{byby}
  L_{m^2}^{\min}=&H_m  =L_{m^2}^{\max},\quad \Re(m)\geq1;\\
  L_{m^2}^{\min}\subsetneq&H_m  \subsetneq L_{m^2}^{\max},\quad |\Re(m)|<1.
\end{align}

\subsection{Main results}

Our new results give descriptions of the domains of various
realizations of $L_\alpha$ for  $\alpha\in\cc$. First of all,
we prove that for $|\Re\sqrt\alpha|<1$ the domain of
$ L_\alpha^{\min}$ does not depend on $\alpha$ and coincides with the
{\em minimal 2nd order Sobolev space} \beq \cH_0^2(\rr_+):= \{f\in
\cH^2(\rr_+)\ |\ f(0)=f'(0)=0\},
\label{sobolev}\eeq 
where
\beq  \cH^2(\rr_+):=
\{f\in 
L^2(\rr_+)\ |\  f''\in L^2(\rr_+)\} \label{sobolev1}\eeq
is the {\em (full) 2nd order Sobolev space}.
We also show that \beq \{\alpha\ |\
|\Re\sqrt\alpha|<1\}\ni\alpha\mapsto L_\alpha^{\min}\eeq is a
holomorphic family of closed operators.

We find the constancy of the domain of the minimal operator quite
surprising and interesting. It contrasts with the fact that
$\cD(L_{\alpha}^{\max})$ for $|\Re\sqrt\alpha|<1$ depends on $\alpha$.
Similarly, $ \cD(H_{m})$ for
$|\Re(m)|<1$ depends on $m$.
                        
The holomorphic family $ L_\alpha^{\min}$ for $|\Re\sqrt\alpha|<1$  consists of operators whose spectrum covers the
whole complex plane. Therefore, the usual approach to holomorphic
families of closed operators based on the study of the resolvent is
not available.

We also study $H_m$ for $\Re(m)\geq1$ (which by \eqref{byby} coincides
with $L_{m^2}^{\min}$ and $L_{m^2}^{\max}$).  We prove that
 for $\Re(m)>1$
its domain also coincides with $\cH_0^2(\rr_+)$.
The most unusual situation occurs in the case $\Re(m)=1$. In this case
we show that the domain of $H_m$ is always larger than
$\ch_0^2(\rr_+)$ and depends on $m$.

Specifying to real $\alpha$, the main result of our paper can be
summarized as follows: Let $L_\alpha^{\min}$ be the closure in
$L^2(\rr_+)$ of the operator
$-\partial_x^2+\frac{\alpha-\frac14}{x^2}$ with domain
$C_\c^\infty(\rr_+)$.\\[1mm]
1) If $\alpha<1$ then $L_\alpha^{\min}$ is Hermitian (symmetric) but
not self-adjoint and its domain is $\ch_0^2(\rr_+)$.
\\[1mm]
2) If $\alpha=1$ then $L_\alpha^{\min}$ is self-adjoint and
$\ch_0^2(\rr_+)$ is a dense subspace of infinite codimension of
its domain. \\[1mm]
3) If $\alpha>1$ then $L_\alpha^{\min}$ is self-adjoint with domain
$\ch_0^2(\rr_+)$.

As a side remark, let us mention two open problems about Bessel operators.
\begin{open}
 $ $ 
\begin{enumerate}
\item
Can the holomorphic family $H_m$  be extended beyond $\Re(m)>-1$?
(Probably not).
\item
Can the holomorphic family $L_\alpha^{\min}$ (hence also $L_\alpha^{\max}$)  be extended beyond
$|\Re\sqrt\alpha|<1$? (Probably not).
\end{enumerate}
\end{open}
Question 1 has already been mentioned in \cite{Derezinski2}.
We hope that both questions can be answered by methods of \cite{DW}.

\subsection{Bilinear Bessel forms}

With  every operator $T$ on a Hilbert space $\cH$ one can
associate the sesquilinear form
\beq (f|Tg),\qquad f,g\in\cD(T).\label{form1}\eeq
One can try to extend \eqref{form1} to a larger domain.
If $T$ is self-adjoint, there is a natural extension to the so-called  {\em form
  domain of $T$}, $\cQ(T):=\cD(\sqrt{|T|})$.
Interpreting $T$ as a bounded map from $\cQ(T)$ to its
anti-dual,  we obtain the sesquilinear form
\beq (f|Tg),\qquad f,g\in\cQ(T),\label{form2}\eeq
which extends \eqref{form1}.

We would like to have a similar construction for Bessel operators,
including non-self-adjoint ones.
Before we proceed we should realize that identities involving
non-self-adjoint operators do not like complex conjugation. Therefore, instead of
sesquilinear forms it is more natural to use bilinear  forms.

Our  analysis of bilinear Bessel forms is based on the
pair of
formal factorizations of the Bessel operator
\begin{align}\label{fact+}
-\partial_x^2 + \Big{(} m^2 - \frac{1}{4}
\Big{)} \frac{1}{x^2}&=
\Big(\partial_x+\frac{\frac12+ m}{x}\Big) 
\Big(-\partial_x+\frac{\frac12+ m}{x}\Big)\\
&=
\Big(\partial_x+\frac{\frac12-m}{x}\Big) 
\Big(-\partial_x+\frac{\frac12- m}{x}\Big).
\label{fact-}\end{align}
In Theorems \ref{facti} and \ref{facti1}
for each $\Re(m)>-1$ we interpret \eqref{fact+} and \eqref{fact-}
as factorizations of the
 Bessel operator  $H_m$ into two closed 1st order operators. They define  natural  bilinear
 forms, which we call {\em Bessel forms}.
For each $\Re(m)>-1$ the corresponding Bessel form is unique, except for $\Re(m)=0$, $m\neq0$, when the two factorizations
 yield two distinct Bessel forms.

Instead of  $\cH_0^2(\rr_+)$, 
the major role is now played by the {\em minimal 1st order Sobolev
space}
\begin{align} \label{sobolev1}\cH_0^1(\rr_+):=\{f\in 
  \cH^1(\rr_+)\ |\ f(0)=0\},
\end{align}
subspace of the {\em (full) 1st order Sobolev space}
\begin{align} \label{sobolev1}\cH^1(\rr_+):=\{f\in 
  L^2(\rr_+)\ |\ f'\in L^2(\rr_+)\}.
\end{align}
Note that $\cH_0^1(\rr_+)$ is the domain of Bessel forms for $\Re(m)>0$.

The analysis of Bessel forms and their factorizations shows a variety
of behaviors depending on the parameter $m$. In particular, there is a
kind of a phase transition  at $\Re(m)=0$.
Curiously, in the analysis of the domain of Bessel operators the
phase transition occurs elsewhere:
at $\Re(m)=1$.

\subsection{Comparison with literature}

The fact that $\cD(L_{\alpha}^{\min})$ does not depend on $\alpha$ for
real $\alpha\in [0,1[$ was first proven in \cite{AlexeevaAnanieva},
see also \cite{AnanievaBudika,AnanievaBudika2}. Actually, the
arguments of \cite{AlexeevaAnanieva} are enough to extend the result
to complex $\alpha$ such that $|\alpha-\frac14|<\frac34$.  The proof
is based on the bound $\|Q\|= \frac34$ of the operator $Q$ on
$L^2(\rr_+)$ given by the integral kernel \beq
Q(x,y)=\frac{1}{x^2}(x-y)\theta(x-y),\label{qu}\eeq where $\theta$ is
the Heaviside function. Our proof is quite similar.  Instead of
\eqref{qu} we consider for $|\Re(m)|<1$ the operator $Q_{m^2}$ with
the kernel \beq
Q_{m^2}(x,y)=\frac{1}{2mx^2}(x^{\frac12+m}y^{\frac12-m}-x^{\frac12-m}y^{\frac12+m})\theta(x-y).\label{qum}\eeq
Note that $Q_\frac14$ coincides with \eqref{qu}.  We prove that the
norm of $Q_{m^2}$ is the inverse of the distance of $m^2$ to the
parabola \eqref{para}.  A simple generalization of the Kato-Rellich
Theorem to closed operators implies then our result about
$\cD(L_\alpha^{\min})$.

In the paper \cite{Derezinski2} on page 567 it is written ``If
$m\neq 1/2$ then $\cD(L^{\min} _m )\neq \ch_0^2 $.'' (In that paper
$L_{m^2}^{\min}$ was denoted $L_{m}^{\min}$). This sentence was not
formulated as a proposition, and no proof was provided. Anyway, in
view of the results of \cite{AnanievaBudika} and of this paper, this
sentence was wrong.

The analysis of Bessel forms in the self-adjoint case, that is for
real $m>-1$,  is well known--it is essentially equivalent to the famous
{\em Hardy inequality}. This subject is discussed e.g. in
  the monograph
\cite{BEL} and in a recent interesting paper \cite{GPS} about a
refinement of the 1-dimensional Hardy's inequality. The latter  paper
contains in particular
many references
about
factorizations of  Bessel operators
in the self-adjoint case.

Results about Bessel forms and their factorizations for complex
parameters are borrowed to a large extent from \cite{Derezinski2}. We
include them in this paper, because they provide an interesting complement to
the analysis of domains of Bessel operators.

\section{Basic closed realizations of the Bessel operator}
\protect\setcounter{equation}{0}

The main topic of this  preliminary section are  closed homogeneous realizations of
$L_\alpha$. We recall their definitions following \cite{Derezinski2, Derezinski1}.

We will denote by $\rr_+$ the open positive half-line, that is
$]0,\infty[$. We will use $L^2(\rr_+)$ as our basic Hilbert space.
We define $L_\alpha^{\max}$ to be the operator given by the
expression $L_\alpha$ with the domain 
\begin{equation*}
\cD(L_\alpha ^{\max}) = \{f\in L^2(\rr_+)\mid L_\alpha  f\in
L^2(\rr_+)\}.
\end{equation*}
We also set $L_\alpha ^{\min}$ to be the 
 closure of the restriction of $L_\alpha^{\max}$ to
$C_{\rm c}^\infty(\rr_+)$.

We will often write $m$ for one of the square roots of $\alpha$, that
is, $\alpha=m^2$.
It is easy to see that  the space of solutions of the differential equation
\begin{equation}
  L_\alpha f=0
\end{equation}
is spanned for $\alpha\neq0$ by
$x^{\frac12+m}$, $x^{\frac12-m}$,
and for $\alpha=0$ by $x^{\frac12}$,  $x^{\frac12}\log x$. 
Note that both solutions are square integrable near $0$ iff $|\Re(m)|<1$.
This is used in \cite{Derezinski2} to show that
   we have 
\begin{align} 
  \cD(L_\alpha^{\max})&=\cD(  L_\alpha^{\min})+
  \cc x^{\frac12+m}\xi+\cc  x^{\frac12-m}\xi, 
  & |\Re\sqrt\alpha|<1,\
  \alpha\neq0;\label{expli2}\\
  \cD(  L_0^{\max}) &=\cD( L_0^{\min})+\cc x^{\frac12}\xi+
  \cc  x^{\frac12}\log(x)\xi ,& \alpha=0;\label{expli3}\\
   \cD( L_\alpha^{\max})&=\cD( L_\alpha^{\min}),&
  |\Re\sqrt\alpha|\geq1. \label{expli4}
\end{align}
Above (and throughout the paper) $\xi$ is any
$C_\mathrm{c}^\infty[0,\infty[$ function such that $\xi=1$ near $0$.

Following \cite{Derezinski2}, for $\Re(m)>-1$ we also introduce another family of closed
realizations of Bessel operators: the operators $H_m$ defined as the
restrictions of $L_{m^2}^{\max}$ to
\beq \label{eq:rem2}
\cD(H_m):=\cD(L_{m^2} ^{\min})
+ \cc x^{\frac12+m}\xi.
\eeq

We will use various basic concepts and facts about 1-dimensional
Schr\"odinger operators with complex potentials. We will use \cite{DG}
as the main reference, but clearly most of them belong to the
well-known folklore.  In particular, we will use two kinds of Green's
operators.  Let us recall this concept, following \cite{DG}.  Let
$L_\mathrm{c}^1(\rr_+)$ be the set of integrable functions of compact
support in $\rr_+$. We will say that an operator
$G: L_\mathrm{c}^1(\rr_+)\to AC^1(\rr_+)$ is a {\em Green's operator
  of $L_\alpha$} if for any $g\in L_\mathrm{c}^1(\rr_+)$
\beq\label{forwar} L_\alpha G g=g.\eeq

\section{The forward Green's operator}
\protect\setcounter{equation}{0}

Let us introduce the operator $G_\alpha^\to$ defined by the kernel
\begin{align}
  G_\alpha^\to(x,y)&:=\frac1{2m}\big(
  x^{\frac12+m}y^{\frac12-m}-x^{\frac12-m}y^{\frac12+m}\big)\theta(x-y), 
  &\alpha\neq0;\\
  G_0^\to(x,y)&:=
  x^{\frac12}y^{\frac12}\log\Big(\frac{x}{y}\Big)\theta(x-y),&\alpha=0.
\end{align}
Note that $G_\alpha^\to$ is a Green's operator in the sense of
\eqref{forwar}. Besides,
\begin{align}
  \supp G_\alpha^\to g\subset\supp
  g+\rr_+,
\end{align}
which is why it is sometimes called the {\em forward
 Green's operator}.

Unfortunately, the operator $G_\alpha^\to$ is unbounded on
$L^2(\rr_+)$.  To make it bounded, for any $a>0$ we can compress it to
the finite interval $[0,a]$, by introducing the operator
$G_\alpha^{a\to}$ with the kernel \beq\label{compre}
G_\alpha^{a\to}(x,y):=\one_{[0,a]}(x)
G_\alpha^\to(x,y)\one_{[0,a]}(y).\eeq It is also convenient to
consider the operator $L_\alpha$ restricted to $[0,a]$. One of its
closed realizations, is defined by the zero boundary condition at $0$
and no boundary conditions at $a$ (see \cite{DG} Def. 4.14). It will
be denoted $L_{\alpha,0}^a$.  By Prop. 7.3 of \cite{DG} we have
$G_\alpha^{a\to}=(L_{\alpha,0}^{a})^{-1}$, and hence \beq \label{koko}
\cD(L_{\alpha,0}^{a})=G_\alpha^{a\to}L^2[0,a].\eeq Now we can describe
the domain of $L_\alpha^{\min}$ with the help of the forward Green's
operator.
\begin{proposition}\label{prio}
  Suppose that $f\in \cD(L_\alpha^{\max})$. Then the following
  statements are equivalent:
  \begin{enumerate}
  \item $f\in \cD(L_\alpha^{\min})$.
  \item For some $a>0$ and $g^a\in L^2[0,a]$ we have
    $f\Big|_{[0,a]}=G_\alpha^\to g^a\Big|_{[0,a]}$.
  \item For all $a>0$ there exists $g^a\in L^2[0,a]$ such that
    $f\Big|_{[0,a]}=G_\alpha^\to g^a\Big|_{[0,a]}$.
  \end{enumerate}
\end{proposition}

\proof The boundary space (\cite{DG} Def. 5.2) of $L_\alpha$ is
trivial at $\infty$ (see \cite{DG} Prop. 5.15). Therefore, for any
$a>0$ we have \beq f\in\cD(L_\alpha^{\min})\ \Leftrightarrow\
f\Big|_{[0,a]} \in\cD(L_{\alpha,0}^a) .\eeq Hence it is enough to
apply \eqref{koko}.  \qed

Define the operator $Q_\alpha:=\frac1{x^2}G_\alpha^\to.$ Its integral
kernel is
\begin{align}
  Q_{\alpha}(x,y)&
=\frac{1}{2m}(x^{-\frac32+m}y^{\frac12-m}-x^{-\frac32-m}y^{\frac12+m})\theta(x-y),\label{qum1} &\alpha\neq0;\\ 
  Q_0(x,y)&:=
 x^{-\frac32}y^{\frac12}\log\Big(\frac{x}{y}\Big)\theta(x-y),&\alpha=0.
\end{align}

\begin{proposition}\label{fou2a}
Assume that $|\Re\sqrt\alpha|<1$.  Then the operator $Q_\alpha$ is
bounded  on $L^2(\rr_+)$, and
\beq\label{fou2}
\|Q_\alpha\|=
\frac{1}{\dist\big(\alpha,(1+\i\rr)^2\big)}\eeq
\end{proposition}

\proof Introduce the unitary operator $U:L^2(\rr_+)\to L^2(\rr)$ given
by \beq\label{dila} (Uf)(t):=\e^{\frac{t}{2}}f(\e^t).\eeq Note that if
an operator $K$ has the kernel $K(x,y)$, then $UKU^{-1}$, has the
kernel $\e^{\frac{t}{2}}K(\e^t,\e^s)\e^{\frac{s}{2}}$. Therefore,
for any $\alpha$ the operator $UQ_\alpha U^{-1}$ has the kernel
\begin{align}
  &\frac{1}{2m}(\e^{-(t-s)(1-m)}-\e^{-(t-s)(1+m)})\theta(t-s),
  &\alpha\neq0;\\ 
  &\e^{-(t-s)}(t-s)\theta(t-s),&\alpha=0.
\end{align}
Thus, it is the convolution by the function
\begin{align}\label{fou}
t\to &\frac{1}{2m}(\e^{-t(1-m)}-\e^{-t(1+m)})\theta(t),&\alpha\neq0;\\
t\to &\e^{-t}t\theta(t),&\alpha=0.\label{fou1a}
\end{align}

Assume now that $|\Re\sqrt\alpha|<1$. Then the function \eqref{fou} is
integrable and we can apply the Fourier transformation defined
  by $(\cF u)(\omega)=(2\pi)^{-1/2}\int\e^{-i\omega t}u(t)\,\d t$.
After this transformation the operator $UQ_\alpha U^{-1}$
becomes the multiplication wrt the Fourier transform of \eqref{fou} or
\eqref{fou1a}, that is \beq\label{fou1} \omega\mapsto
\frac1{(1+\i\omega)^2-m^2}.  \eeq Thus the norm of
$UQ_\alpha U^{-1}$, and hence also of $Q_\alpha$, is the supremum of
the absolute value of \eqref{fou1}.  \qed

\begin{remark}
  The operator $Q_\alpha$ belongs to the class of operators analyzed
  in \cite{Stein} on
  p. 271,
  which goes  back to Hardy-Littlewood-Polya \cite{HLP}
   p. 229. 
\end{remark}

Proposition \eqref{fou2a} for $\alpha=\frac14$ is especially important
and simple.  This case was noted in cf.\ \cite[p.\ 566]{Derezinski2}
and \cite[Lemma 2.2]{AlexeevaAnanieva}. It can be written as
\begin{align}
  &g(x):= x^{-2}\int_0^x (x-y)f(y)\d y \ \Rightarrow\ 
    \|g\|\leq
 \frac{4}{3}\|f\|.  \label{eq:estima}
\end{align}
One can remark that \eqref{eq:estima}  is essentially equivalent to the
  one-dimensional version of the classical  {\em Rellich's
 inequality}, see e.g. \cite[(6.1.1)]{BEL}:
\beq
\int_0^\infty\frac{|u|^2}{x^4}\d
x\leq\frac{16}{9}\int_0^\infty|u''|^2\d x,\quad u\in C_\mathrm{c}^\infty(\rr_+),\eeq
where we identify $f=u''$ and $g=\frac{u}{x^2}$.

The proof of the following proposition uses only the simple estimate
\eqref{eq:estima}.
\begin{proposition}\label{pr:domains}
$\cD(L_\alpha^{\max})\cap \cD(L_\beta^{\max})=\ch_0^2(\rr_+)$ if
$\alpha\neq\beta$. 
\end{proposition}

 \proof We have $f\in\cD(L_\alpha ^{\max})$ if and only if
  $f\in L^2(\rr_+)$ and $-f''+(\alpha-1/4)x^{-2}f\in L^2(\rr_+)$ hence
  if we also have $f\in\cD(L_\beta^{\max})$ then
  $(\alpha-\beta)x^{-2}f\in L^2(\rr_+)$ and since $\alpha\neq\beta$ we
  get $x^{-2}f\in L^2(\rr_+)$ hence $f''\in L^2(\rr_+)$. Recall that
  $f,f''\in L^2(\rr_+)$ implies $f\in\ch^1(\rr_+)$ and
  $\|f'\|^2_{L^2(\rr_+)}\leq\|f\|^2_{L^2(\rr_+)}\|f''\|^2_{L^2(\rr_+)}$.
  It follows that $f$ is absolutely continuous and
  $f(x)=a+\int_0^x f'(y)\d y$ for some constant $a$ and $f'$ is
  absolutely continuous and $f'(x)= b+\int_0^x f''(y)\d y$ for some
  constant $b$, thus
\[
  f(x)=a+bx+\int_0^x\int_0^y f''(z)\d z \d y
  =a+bx+x^2g(x),\qquad g(x):=x^{-2}\int_0^x (x-y)f''(y)\d y.
\]
Then, by \eqref{eq:estima}
\begin{align}
  &\|g\|_{L^2(\rr_+)}\leq\frac{4}{3}\|f''\|_{L^2(\rr_+)}.  \label{eq:estim}
\end{align}
Thus $x^{-2}f(x)= ax^{-2}+bx^{-1}+g(x)$ where $g\in L^2(\rr_+)$, so
$\int_0^1|x^{-2}f(x)|^2\d x<\infty$ if and only if $a=b=0$, so that
$f(x)=\int_0^x (x-y)f''(y)\d y$ and $f'(x)=\int_0^x f''(y)\d y$, hence
$f\in \ch_0^2(\rr_+)$.

Reciprocally, if $f\in \ch_0^2(\rr_+)$ then $x^{-2}f\in L^2(\rr_+)$
with $\|x^{-2}f\|_{L^2(\rr_+)}\leq\frac{4}{3}\|f''\|_{L^2(\rr_+)}$ by
\eqref{eq:estima}, hence $f\in\cD(L_\alpha^{\max})$ for all $\alpha$.
\qed

\section{Domain of Bessel operators for $|\Re(m)|<1$}    
\protect\setcounter{equation}{0}

Below we state the first main result of our paper (which is an
extension of a result of \cite{AlexeevaAnanieva}).

\bet \label{alexeeva}
If $|\Re\sqrt\alpha|<1$, then $\cD(L_\alpha^{\min})=\ch_0^2(\rr_+)$.
Moreover,
\beq\label{fou4}
\left\{\alpha\in\cc\ |\ |\Re\sqrt\alpha|<1\right\}\ni\alpha\mapsto 
L_\alpha^{\min}\eeq 
is a holomorphic family of closed operators. \eet

The proof of this theorem is based on the following lemma.

\begin{lemma}\label{kato}
  Let $|\Re\sqrt\alpha|<1$ and $f\in\cD(L_\alpha^{\min})$. Then
  \beq
  \|x^{-2}f\|\leq \frac{1}{\dist\big(\alpha,(1+\i\rr)^2\big)}\|L_\alpha^{\min}f\|.
  \eeq
\end{lemma}

\proof Let $a>0$. Set $g:=L_\alpha^{\min}f$, $f^a:=f\Big|_{[0,a]}$,
$g^a:=g\Big|_{[0,a]}$. Let $G_\alpha^{a\to}$ be as in \eqref{compre}.
As in the proof of Prop. \ref{prio},
\beq f^a=G_\alpha^{a\to}g^a.\eeq
So
\begin{align}
  \|x^{-2}f\|&=\lim_{a\to\infty}\|x^{-2}f^a\|
  \\=\lim_{a\to\infty}\|x^{-2}G_\alpha^{a\to}g^a\|&=\|Q_\alpha g\|
\leq \frac{1}{\dist\big(\alpha,(1+\i\rr)^2\big)} \|g\|. 
\quad\Box
\end{align}

\noindent{\em Proof of Theorem \ref{alexeeva}.}
We can cover the region on the lhs of \eqref{fou4} by disks touching
the boundary of this region, that is, \eqref{para}. Inside each disk
we apply Thm \ref{kato-rellich} 
and
Lemma \ref{kato}.
We obtain in particular, that
if $|\Re\sqrt\alpha_i|<1$, $i=1,2$,
then $\cD(L_{\alpha_1}^{\min})= \cD(L_{\alpha_2}^{\min})$.
But clearly $\cD(L_{\frac14}^{\min})=\ch_0^2(\rr_+)$. \qed

\bet\label{qqaqq1}
We have
\begin{align} 
  \cD(L_\alpha^{\max})&=\cH_0^2+
  \cc x^{\frac12+m}\xi+\cc  x^{\frac12-m}\xi, 
  & |\Re\sqrt\alpha|<1,\
  \alpha\neq0;\label{expli2a}\\
  \cD(  L_\alpha^{\max}) &=\cH_0^2+\cc x^{\frac12}\xi+
  \cc  x^{\frac12}\log(x)\xi ,& \alpha=0.\label{expli3a}
\end{align}
Besides,\beq\label{qqaqq}
\cD(L_{\alpha_1}^{\max})\neq \cD(L_{\alpha_2}^{\max}), \qquad
\alpha_1\neq\alpha_2, \quad|\Re\sqrt\alpha_i|<1,\quad
i=1,2.\eeq
Furthermore,
\beq\label{fou5}
\left\{\alpha\in\cc\ |\ |\Re\sqrt\alpha|<1\right\}\ni\alpha\mapsto 
L_\alpha^{\max}\eeq 
is a holomorphic family of closed operators.
\eet

\proof Using $ \cD(L_\alpha^{\min})=\cH_0^2$, 
\eqref{expli2} and 
\eqref{expli3} can be now rewritten as
\eqref{expli2a} and 
\eqref{expli3a}.

Clearly,  $x^{\frac12+m}\xi$ and
$x^{\frac12}\log(x)\xi$ do not belong to
$\cH_0^2(\rr_+)$ (because their second derivatives 
are not square integrable).
Therefore, $ \cD(L_\alpha^{\max})\neq\cH_0^2(\rr_+)$.
This together with Proposition \ref{pr:domains} implies
\eqref{qqaqq}.

We have $(L_\alpha^{\min})^*=L_{\bar\alpha}^{\max}$.
Therefore, to obtain the holomorphy we can use Proposition \ref{adjo}.
\qed

The most important holomorphic family of Bessel operators is
\beq\label{holom} \{m\in\cc\ |\ \Re(m)>-1\}\ni m\mapsto H_m.\eeq
Its holomorphy has been proven in 
\cite{Derezinski2}.
Using  arguments similar to those in the proof of
Theorem \ref{qqaqq1} we obtain a closer description of this family in
the region $|\Re(m)|<1$.

\bet \label{holom1}
We have
\begin{align} 
  \cD(H_m)&=\cH_0^2+
  \cc x^{\frac12+m}\xi, 
  & |\Re(m)|<1.\end{align}
Besides,
if $m_1\neq m_2$ and $|\Re(m_i)|<1$, $i=1,2$, then $ \cD(H_{m_1})\neq \cD(H_{m_2})$.
\eet

\section{Two-sided Green's operator}\label{s:tsgo}
\protect\setcounter{equation}{0}

For any $m\in\cc$, $m\neq0$, let us introduce the operator $G_m$ with the kernel
\begin{align}\label{twos}
G_m(x,y)&:=\frac1{2m}\left(x^{\frac12+m}y^{\frac12-m}\theta(y-x)+
          x^{\frac12-m}y^{\frac12+m}\theta(x-y)\right).
\end{align}
Recall that $\theta$ is the Heaviside function. \eqref{twos} is one of Green's
operators of $L_{m^2}$ in the sense of \eqref{forwar}, Following
\cite{DG}, we will call it  the {\em two-sided Green's operator}.

The operator $G_m$ is not bounded on $L^2(\rr_+)$ for any
$m\in\cc$. However, at least for
$\Re(m)>-1$, it is useful in the $L^2$ setting.
\begin{proposition}\label{cond0}
  Let $\Re(m)>-1$, $m\neq0$ and $a>0$.
  \begin{enumerate}\item
  If $g\in L^2[0,a]$,
  then
    \beq
    f(x)=G_mg(x)=\int_0^\infty G_m(x,y)g(y)\d y\eeq
    is well defined, belongs to $
    \in AC^1]0,\infty[$ and $L_\alpha f=g$.
  \item Conversely, if $f\in AC^1]0,\infty[$,
    $L_\alpha f=g\in L^2[0,a]$, then there exist $c_+,c_-$ such that
    \begin{align} f(x)&=c_+x^{\frac12+m}+c_-x^{\frac12-m}+G_mg(x),\quad
      m\neq0.      \end{align}
\end{enumerate}
\end{proposition}

  \proof
 Note first that $\Re(m)>1$ implies $x^{\frac12\pm m}$ is
  locally in $L^2$. Using this,
  the proof of the first part of the proposition is a
straightforward computation done, in a more general setting, in
\cite{DG}, see \S2.7 and Definition 2.10 there. For the second part,
note that $L_\alpha(f-G_mg)=0$ by the first part of the proposition,
and that the two functions $x^{\frac12\pm m}$ give a basis of
 the nullspace of
$L_\alpha$.  \qed

Let us introduce the
operator $Z_m:=\frac{1}{x^2}G_m$ 
with the kernel
\begin{align}
Z_m(x,y)&=\frac{1}{2m}\left(x^{-\frac32+m}y^{\frac12-m}\theta(y-x) 
          +x^{-\frac32-m}y^{\frac12+m}\theta(x-y) \right).
\end{align}

\begin{proposition} \label{zet}
  Let $\Re(m)>1$. Then $Z_m$ is bounded and
  \begin{equation}\label{lha}
    \|Z_m\|=\frac{1}{\dist\big(m^2,(1+\i\rr)^2\big)}
    \end{equation}
\end{proposition}

\proof
If $U$ is given by \eqref{dila}, then $UZ_mU^{-1}$ has the kernel
\begin{align}
\frac{1}{2m}\left(\e^{-(m-1)(s-t)}\theta(s-t)
+\e^{-(m+1)(s-t)}\theta(t-s)\right).  
\end{align} 
If $\Re(m)>1$, after the Fourier transformation (defined as
  in the proof of Proposition \ref{fou2a}) it becomes the
  multiplication by the function
\begin{align}
  \omega\mapsto \frac1{2m}\Big(\frac{1}{(m-1-\i\omega)}+
  \frac{1}{1+m+\i\omega)}\Big)=\frac{1}{m^2-(1+\i\omega)^2},
\end{align}
whose supremum is the right hand side of \eqref{lha}. \qed


\section{Domain of Bessel operators for
  $\Re(m)>1$}\label{s:dbo} 
\protect\setcounter{equation}{0}

For $\Re(m)\geq 1$ there is a unique closed Bessel operator.  We will
see in the following theorem that its domain is again the minimal 2nd
order Sobolev space, except at the boundary $\Re(m)=1$, cf.\ Section
\ref{s:dbo1}.

\bet \label{th:mge1}
Let $\Re(m)>1$. Then $\cD(H_m)=\ch_0^2(\rr_+)$.
\eet

\proof We know that $\ch_0^2(\rr_+)\subset \cD(L_{m^2}^{\max})$ for
any $m$. But for $\Re(m)>1$ we have $L_{m^2}^{\max}=H_m$. This proves
the inclusion $\ch_0^2(\rr_+)\subset \cD(H_{m})$.

Let us prove the converse inclusion. Let $f\in\cD(H_m)$. It is enough
to assume that $f\in L^2[0,1]$. Let $g:=H_m f$. Then $g\in
L^2[0,1]$. By Prop. \ref{cond0}, we can
write 
\beq\label{eq:msol}
f(x)=c_+x^{\frac12+m}+c_-x^{\frac12-m}+\frac{x^{\frac12+m}}{2m}
\int_x^1y^{\frac12-m}g(y)\d y +\frac{x^{\frac12-m}}{2m}
\int_0^xy^{\frac12+m}g(y)\d y.  \eeq For $x>1$ we have \beq
f(x)=c_+x^{\frac12+m}+x^{\frac12-m}\left(c_- +\frac{1}{2m}
  \int_0^1y^{\frac12+m}g(y)\d y\right), \eeq hence $c_+=0$.  We have,
for $x\to0$,
\begin{align}\label{condi4}
\left|    x^{\frac12+m}
\int_x^1y^{\frac12-m}g(y)\d y \right|&\leq x\int_0^1|g(x)|\d
                                          y\to0;\\\label{condi5}
    \left| x^{\frac12-m}\int_0^xy^{\frac12+m}g(y)\right|\d y&\leq
     x\int_0^x|g(y)|                                                          \d y\to0.
\end{align}
$x^{\frac12-m}$ is not square integrable near zero. Hence $c_-=0$.  Thus
\beq f(x)=\frac{x^{\frac12+m}}{2m} \int_x^1y^{\frac12-m}g(y)\d y
+\frac{x^{\frac12-m}}{2m} \int_0^xy^{\frac12+m}g(y)\d y.  \eeq
By \eqref{condi4} and \eqref{condi5}, $\lim\limits_{x\to0}f(x)=0$. Now
\begin{align}
f'(x)=\frac{(\frac12+m)x^{-\frac12+m}}{2m}
  &\int_x^1y^{\frac12-m}g(y)\d y +
    \frac{(\frac12-m)x^{-\frac12-m}}{2m}
\int_0^xy^{\frac12+m}g(y)\d y,\\
\left|x^{-\frac12-m}\int_0^xy^{\frac12+m}g(y)\d y \right|&\leq
\int_0^x|g(y)|\d y\to 0,\\ \label{condi6} 
  \left|x^{-\frac12+m} \int_x^1y^{\frac12-m}g(y)\d y \right|&\leq
\int_0^\epsilon|g(y)|\d y
+x^{-\frac12+\Re(m)}\int_\epsilon^1y^{\frac12-\Re(m)}|g(y)|\d y.
\end{align}
For any $\epsilon>0$, the second term on the right of \eqref{condi6}
goes to zero. The first, by making $\epsilon$ small, can be made
arbitrarily small. Therefore \eqref{condi6} goes to zero.  Thus
$\lim\limits_{x\to0}f'(0)=0$.

Finally  
\begin{align}
  f''(x)+g(x)=&\frac{(m^2-\frac14)x^{-\frac32+m}}{2m}
          \int_x^1y^{\frac12-m}g(y)\d y 
          +\frac{(m^2-\frac14)x^{-\frac32-m}}{2m}
                \int_0^xy^{\frac12+m}g(y)\d y\\
  =&\Big(m^2-\frac14\Big)Z_mg(x).
\end{align}
By Proposition \ref{zet} $Z_m$ is bounded. Hence $f''\in
L^2(\rr_+)$. Therefore, $f\in \ch_0^2(\rr_+)$. \qed

\section{Domain of Bessel operators for $\Re(m)=1$}\label{s:dbo1}
\protect\setcounter{equation}{0}

In this section we treat the most complicated situation concerning the
domain of $H_m$, namely the case $\Re(m)=1$.  By \eqref{byby} we
then have $H_m=L_{m^2}^{\min}=L_{m^2}^{\max}$. We
prove the following theorem.

\begin{theoreme}\label{th:rem11}
  Let $\Re(m)=1$.
  \begin{enumerate}
  \item $\ch_0^2(\rr_+)$ is a dense subspace of $\cD(H_m)$
    of infinite codimension.
  \item If $\xi$ is a $C_\mathrm{c}^2[0,\infty[$ function equal $1$
    near zero, then $x^{\frac12+m}\xi\in \cD(H_m)$ but
    $x^{\frac12+m}\xi\not\in\ch_0^2(\rr_+)$.
  \item If $\Re(m')=1$ and $m\neq m'$, then
    $\cD(H_m)\cap\cD(H_{m'})=\ch_0^2(\rr_+)$.
\end{enumerate}
\end{theoreme}

By \eqref{byby}, it is clear that $\ch_0^2(\rr_+)\subset\cD(H_m)$ and
$x^{\frac12+m}\xi\in \cD(H_m)$. The density of $\ch_0^2(\rr_+)$ in
$\cD(H_m)$ is a consequence of $H_m=L_{m^2}^{\min}$.  The last
assertion of the theorem is a special case of Proposition
\ref{pr:domains}.  In the rest of this section we construct an
infinite dimensional vector subspace  $\cV$ of $\cD(H_m)$ such that
$\cV\cap\big(\ch_0^2(\rr_+)+\cc x^{\frac12+m}\xi\big)=\{0\}$,
which will finish the proof of the theorem.

Let us study the behaviour at zero of the functions in
$\cD(H_m)$. For functions in the subspace $\ch_0^2(\rr_+)+\cc x^{\frac12+m}\xi$ this
is easy, cf.\ the next lemma, but this is not so trivial for the other
functions.

\begin{lemma}\label{lm:limo}
  If $f=f_0 + cx^{\frac12+m}\xi\in\ch_0^2(\rr_+)+\cc x^{\frac12+m}\xi$ then
\begin{equation}\label{eq:limo}
  c=\lim_{x\to0} x^{-\frac12-m}f(x). 
\end{equation}
\end{lemma}

\proof If $f_0\in\ch_0^2(\rr_+)$ then
$f_0(x)=\int\limits_0^x(x-y)f_0''(y)\d y$. Therefore,
$\sqrt{3} |f_0(x)| \leq x^{\frac32}\|f_0''\|_{L^2[0,x]}$ and since
$\Re(m+\frac12)=\frac32$ we get
$\lim\limits_{x\to0} x^{-m-\frac12}f_0(x)=0$, which implies
\eqref{eq:limo}.  \qed

Let $a>0$. Let $G_m^a$ be the operator $G_m$ compressed to the
interval $[0,a]$. Its kernel is
\beq\label{gma}
G_m^a(x,y)=\one_{[0,a]}(x)G_m(x,y)\one_{[0,a]}(y).
\eeq
We will write $L_\alpha^{a,\max}$ for the maximal realization of operator $L_\alpha$ on $L^2[0,a]$.

\begin{lemma} \label{lm:inj}
  Let $\Re(m)>-1$. Then $G_m^a$ is a bounded operator on
  $L^2[0,a]$. If $g\in   L^2[0,a]$, then $G_m^ag\in\cD(
  L_{m^2}^{a,\max})$
  and $L_{m^2}^{a,\max} G_m^ag=g$.
  Consequently, $G_m^a$ is injective. \end{lemma}

\proof We check that 
\eqref{gma} belongs to $L^2\big([0,a]\times[0,a]\big)$. This proves that $G_m^a$ is Hilbert
Schmidt, hence bounded.
$G_m^a$ is a right inverse of $L_{m^2}^{a,\max}$,  because
  $G_m$ is a right inverse of $L_{m^2}$ (see Proposition \ref{cond0}). \qed

\begin{lemma} \label{lm:zeroint}
Let $\Re(m)=1$. Let $g\in L^2[0,a]$ and $f=G_m^ag$. Then
\begin{equation}\label{eq:ellr1}
  \lim_{x\to0}\left(2mx^{-\frac12-m}f(x) -
    \int_x^ay^{\frac12-m}g(y)\d y \right)=0. 
\end{equation}
Therefore, if
\beq\label{eq:limi}
\lim_{x\to0}\int_x^ay^{\frac12-m}g(y)\d y\eeq
does not exist, then $f=G_m^ag
\not\in\ch_0^2(\rr_+)+\cc x^{\frac12+m}\xi$.
\end{lemma}

\proof We have
\begin{equation}
  2mx^{-\frac12-m}f(x) =\int_x^ay^{-\frac12-m}g(y)\d y
  +x^{-2m}\int_0^xy^{\frac12+m}g(y)\d y.
\end{equation}
Since $\Re(m)=1$ the absolute value of the second term on the
right hand side is less than
\[
 x^{-\frac12}\int_0^x(y/x)^{\frac32}|g(y)|\d y
  \leq x^{-\frac12}\int_0^x|g(y)|\d y\leq \|g\|_{L^2[0,x]}
\]
This proves \eqref{eq:ellr1}.
If $f=G_m^ag \in\ch_0^2(\rr_+)+\cc x^{\frac12+m}\xi$, then by
\eqref{eq:ellr1} and \eqref{eq:limo} there exists
\eqref{eq:limi}. This proves the second statement of the lemma. \qed

\begin{lemma}
  Let $\Re(m)=1$. There exists an infinite dimensional subspace
  $\cV\subset \cD(H_m)$ such that
  \beq \label{eq:cfh}
\cV\cap\big(\ch_0^2(\rr_+)+\cc
  x^{\frac12+m}\xi\big)=\{0\}.
\eeq
\end{lemma}


\proof For each $\tau\in]\frac12,1[$ let $g_\tau\in C^2(]0,1])$,
for $0<x<\frac12$ given by
\[
g_\tau(x)= x^{-\frac32+m} \big(\ln(1/x)\big)^{-\tau}
\]
and arbitrary on $[\frac12,1]$.  Then for $x<\frac12$ we have
\[
  |g_\tau(x)|^2= x^{-1} \big(\ln(1/x)\big)^{-2\tau}=
   (2\tau-1)^{-1}\frac{\d}{\d x} \big(\ln(1/x)\big)^{1-2\tau}.
\]
Hence
\[
\int_0^{\frac12} |g_\tau(x)|^2\d x =(2\tau-1)^{-1}(\ln2)^{1-2\tau},
\]
and $g_\tau\in L^2[0,1]$.
Moreover, if $x<\frac12$ 
then
\[
  x^{\frac12-m}g_\tau(x)= x^{-1} \big(\ln(1/x)\big)^{-\tau}=
   (\tau-1)^{-1}\frac{\d}{\d x} \big(\ln(1/x)\big)^{1-\tau}.
\]
Hence
\[
  \int_x^{\frac12}y^{-\frac12}g_\tau(y)\d y=(\tau-1)^{-1}(\ln2)^{1-\tau}
  +(1-\tau)^{-1}\big(\ln(1/x)\big)^{1-\tau} \to \infty
  \quad\text{as } x\to0.  
\]
Let $\cG$ be the vector subspace of $L^2[0,1]$ generated by the
functions $g_\tau$ with $\frac12<\tau<1$. Note that each finite
set $\{g_\tau\mid \tau\in A\}$ with $A\subset]\frac12,1[$ finite
is linearly independent. Indeed, if
$\sum\limits_{\tau\in A}c_\tau g_\tau=0$ and $\sigma=\min A$ and
$\tau\neq\sigma$ then
$\frac{g_\tau(x)}{g_\sigma(x)}=\big(\ln(1/x)\big)^{\sigma-\tau}
\to0$ as $x\to0$ so we get $c_\sigma=0$, etc. Moreover, for each not
zero $g=\sum\limits_{\tau\in A}c_\tau g_\tau\in\cG$ (with
$c_\tau\neq0$) we have
$\lim\limits_{x\to0}\left|\int_x^1 y^{-\12}g(y)\d
  y\right|=\infty$.  Indeed, we may assume $c_\sigma=1$ and then,
\begin{align*}
  \int_x^{\frac12}y^{-\frac12}g(y)\d y=&
  (1-\sigma)^{-1} \big(\ln(1/x)\big)^{1-\sigma}\\
  &+\sum\limits_{\tau\in A}c_\tau (\tau-1)^{-1}(\ln2)^{1-\tau}+
\sum_{\tau\neq\sigma}  c_\tau(1-\tau)^{-1}\big(\ln(1/x)\big)^{1-\tau},
\end{align*}
and the first term  on the right hand side tends to $+\infty$ more rapidly
than all the other, hence
\[
\left|\int_x^{\frac12}y^{-\frac12}g(y)\d y\right|\geq \frac{1}{2(1-\sigma)}
\big(\ln(1/x)\big)^{1-\sigma}
\]
if $x$ is small enough. 

 Finally,  let $\varphi\in
    C_\mathrm{c}^\infty[0,\infty[$ equal $1$ on $[0,1]$. Let us define $\cV$ as
  the space of functions
on $\rr_+$
of the form
  $f=\varphi G_mg$ with $g\in\cG$.   By Lemma
  \ref{lm:inj}, $G_m\one_{[0,1]}(x)$ is injective. Hence $\cV$ is infinite dimensional and it satisfies
  \eqref{eq:cfh} by Lemma \ref{lm:zeroint}.  \qed

\section{Bilinear forms associated with Bessel operators}

As noted in the introduction, in this section we will avoid complex
conjugation.  Thus in the place of the usual sesquilinear scalar
product \beq (f|g):=\int_0^\infty \bar{f(x)}g(x)\d x, \eeq we will
prefer to use the bilinear product \beq \langle
f|g\rangle:=\int_0^\infty f(x)g(x)\d x,\label{bili}\eeq Clearly,
\eqref{bili} is well defined for $f,g\in L^2(\rr_+)$.  Instead of the
usual adjoint $T^*$ we will use the transpose $T^\#$, defined with respect to
\eqref{bili}, see \cite{DG}.

An important role will be played by the 1st order operators given by  the formal expression
\beq\label{asas}
A_\rho:=\partial_x-\frac{\rho}{x}.
\eeq
 A detailed analysis of   \eqref{asas} has been done in
 \cite{Derezinski2}, where the 
notation was slightly different: 
$A_\rho:=-\i(\partial_x-\frac{\rho}{x})$. Let us recall the main
points of that analysis.

In the usual way we define two closed realizations of $A_\rho$: the
minimal and the maximal one, denoted $A_\rho^{\min}$, resp.
$A_\rho^{\max}$.  The following theorem was (mostly) proven in
Section 3 of \cite{Derezinski2}.  For the proof
  of the infinite codimensionality assertion in 6 see the proof of
  Lemma 3.9 there (where $\gamma$ is arbitrary $>\frac{1}{2}$).

\bet
\begin{enumerate}
\item
$A_\rho^{\min}\subset A_\rho^{\max}$.
\item 
$A_\rho^{\min\#}=-A_{-\rho}^{\max}$,\quad $A_\rho^{\max\#}=-A_{-\rho}^{\min}$.
\item
$A_\rho^{\min}$ and $ A_\rho^{\max}$ are homogeneous of degree $-1$.
\item
$A_\rho^{\min}= A_\rho^{\max}$ iff $|\Re(\rho)|\geq\frac12$. If this is
the case, we will often denote them simply by $A_\rho$
\item
If $\Re(\rho)\neq\frac12$, then $\cD(A_\rho^{\min})=\cH_0^1$.
\item
If $\Re(\rho)=\frac12$, then $\cH_0^1+\cc x^\rho\xi$ is a dense subspace of
$\cD(A_\rho)$ of infinite codimension.
\item If $|\Re(\rho)|<\frac12$, then $\cD(A_\rho^{\max})=\cH_0^1+\cc
  x^\rho\xi \neq\cH_0^1$. 
\item If $\Re(\rho),\Re(\rho')\in]-\frac12,\frac12]$ and
    $\rho\neq\rho'$ then $\cD(A_\rho^{\max})\neq \cD(A_{\rho'}^{\max})$.
\end{enumerate}
\eet

Now let us describe possible factorizations of $H_m$ into operators of
the form
$A_\rho^{\min}$ and $ A_\rho^{\max}$. On the formal level they
correspond to one of the factorizations \eqref{fact+} and \eqref{fact-}.

\bet\label{fact3}
\begin{enumerate}
\item
  For $\Re(m)>-1$ we have
  \beq\langle f|H_mg\rangle=\langle A_{\frac12+m}^{\max}f|
   A_{\frac12+m}^{\max}g\rangle,\quad
   f\in\cD(A_{\frac12+m}^{\max}),\quad g\in \cD(A_{\frac12+m}^{\max})\cap\cD(H_m).
   \eeq
   Moreover,
   \begin{align}
   \cD(H_m)&=\left\{f\in\cD(A_{\frac12+m}^{\max})\ |\
   A_{\frac12+m}^{\max}f\in\cD(   A_{-\frac12-m}^{\min})\right\},\\
     H_mf&=- A_{-\frac12-m}^{\min}A_{\frac12+m}^{\max} f,\qquad f\in\cD(H_m).
           \end{align}
\item
  For $\Re(m)>0$ we have
  \beq\langle f|H_mg\rangle=\langle A_{\frac12-m}^{\min}f|
   A_{\frac12-m}^{\min}g\rangle,\quad
   f\in\cD(A_{\frac12-m}^{\min}),\quad g\in \cD(A_{\frac12-m}^{\min})\cap\cD(H_m).
   \eeq
   Moreover,
   \begin{align}
   \cD(H_m)&=\left\{f\in\cD(A_{\frac12-m}^{\min})\ |\
   A_{\frac12-m}^{\min}f\in\cD(   A_{-\frac12+m}^{\max})\right\},\\
     H_mf&=- A_{-\frac12+m}^{\max}A_{\frac12-m}^{\min} f,\qquad f\in\cD(H_m).
           \end{align}
         \end{enumerate}
 \label{facti}\eet
 
 The factorizations described in Theorem \ref{facti} can be used to
 define bilinear forms corresponding to $H_m$. For details of
   the proof, we refer again to \cite{Derezinski2}, especially pages
   571--574 and 577.

\bet\label{facti1}
The following bilinear forms are extensions of 
\beq
\langle f|H_mg\rangle=\langle H_mf|g\rangle,\qquad f,g\in\cD(H_m),\eeq
to larger domains:
\begin{enumerate}
\item For $1\leq\Re(m)$,
  \begin{align}
\langle A_{\frac12+m}f|
   A_{\frac12+m}g\rangle&=\langle A_{\frac12-m}f|
    A_{\frac12-m}g\rangle,&\quad f,g\in\cH_0^1.
  \end{align}
\item For $0<\Re(m)<1$,
  \begin{align}
\langle A_{\frac12+m}f|
   A_{\frac12+m}g\rangle&=\langle A_{\frac12-m}^{\min}f|
    A_{\frac12-m}^{\min}g\rangle,&\quad f,g\in\cH_0^1.
  \end{align}
\item For $\Re(m)=0$,
  \begin{align}
\langle A_{\frac12+m}f|
    A_{\frac12+m}g\rangle,&&\quad f,g\in\cD(A_{\frac12+m}) \supset\cH_0^1+\cc x^{\frac12+m}\xi,\\
    \langle A_{\frac12-m}f|
   A_{\frac12-m}g\rangle,&&\quad f,g\in\cD(A_{\frac12-m}) \supset\cH_0^1+\cc x^{\frac12-m}\xi.
  \end{align}
                   \item For $-1<\Re(m)<0$,
  \begin{align}
\langle A_{\frac12+m}^{\max}f|
   A_{\frac12+m}^{\max}g\rangle,&&\quad f,g\in\cH_0^1+\cc x^{\frac12+m}\xi.
  \end{align}

  \end{enumerate}
\eet

Note that for $\Re(m)>0$ both factorizations yield the same quadratic
form. This is no longer true for $\Re(m)=0$, $m\neq0$, when there are
two distinct quadratic forms with distinct domain corresponding to
$H_m$. Finally, for $-1<m<0$, and also for $m=0$, we have a unique
factorization.

Let us finally specialize Theorem 
\ref{facti1} to real $m$. The following theorem is essentially
identical with Thm 4.22 of \cite{Derezinski2}.

\bet\label{facti2}
For real  $-1<m$ the operators $H_m$ are positive and
self-adjoint. The corresponding sesquilinear form
can be factorized as follows:
\begin{enumerate}
\item For $1\leq m$,
  \begin{align}
(\sqrt{H_m}f|\sqrt{H_m}g)= (A_{\frac12+m}f|
   A_{\frac12+m}g)&=( A_{\frac12-m}f|
    A_{\frac12-m}g),&\quad f,g\in\cQ(H_m)=\cH_0^1.
  \end{align}
$H_m$ is essentially self-adjoint on $C_\mathrm{c}^\infty(\rr_+)$.
\item For $0<m<1$,
  \begin{align}(\sqrt{H_m}f|\sqrt{H_m}g)=
( A_{\frac12+m}f|
   A_{\frac12+m}g)&=( A_{\frac12-m}^{\min}f|
    A_{\frac12-m}^{\min}g),&\quad f,g\in\cQ(H_m)=\cH_0^1.
  \end{align}
  $H_m$ is the Friedrichs extension of $L_{m^2}$ restricted to $C_\mathrm{c}^\infty(\rr_+)$.
\item For $m=0$,
  \begin{align}
(\sqrt{H_0}f|\sqrt{H_0}g)= ( A_{\frac12}f|
    A_{\frac12}g),&&\quad f,g\in\cQ(H_0)=\cD(A_{\frac12}) \supsetneq\cH_0^1+\cc
                     x^{\frac12}\xi.
  \end{align}
    $H_0$ is both the Friedrichs and Krein  extension of $L_0$ restricted to $C_\mathrm{c}^\infty(\rr_+)$.
                   \item For $-1<m<0$,
  \begin{align}(\sqrt{H_m}f|\sqrt{H_m}g)=
(A_{\frac12+m}^{\max}f|
   A_{\frac12+m}^{\max}g),&&\quad f,g\in\cQ(H_m)=\cH_0^1+\cc x^{\frac12+m}\xi.
  \end{align}
    $H_m$ is the Krein extension of $L_{m^2}$ restricted to $C_\mathrm{c}^\infty(\rr_+)$.
\end{enumerate}
\eet

\appendix

\section{Holomorphic families of closed operators and the
  Kato-Rellich Theorem}

In this appendix we describe a few general concepts and facts from the
operator theory, which we use in our paper.

The definition (or actually a number of equivalent definitions) of a
{\em holomorphic family of bounded operators} is quite obvious and
does not need to be recalled. In the case of unbounded operators the
situation is more subtle.

Suppose that $\Theta$ is an open subset of $\cc$, $\cH$ is a Banach
space, and $\Theta\ni z\mapsto H(z)$ is a function  whose values
are closed operators on $\cH$.  We say that this is a {\em
  holomorphic family of closed operators} if for each $z_0\in\Theta$
there exists a neighborhood $\Theta_0$ of $z_0$, a Banach space
$\cK$ and a holomorphic family of bounded operators $\Theta_0\ni
z\mapsto A(z)\in B(\cK,\cH)$ such that $\Ran A(z)=\cD(H(z))$ and
\begin{equation*}
\Theta_0\ni z\mapsto H(z)A(z)\in B(\cK,\cH)\label{holo2}
\end{equation*}
is a holomorphic family of bounded operators.

The following theorem is essentially a version of the well-known
Kato-Rellich Theorem generalized from self-adjoint to closed operators:

\bet\label{kato-rellich}
Suppose that $A$ is a closed operator on a Hilbert space $\cH$.  Let
$B$ be an operator $\cD(A)\to \cH$ such that
\beq
\|Bf\|\leq c\|A f\|,\quad f\in\cD(A).\eeq
Then for $|z|<\frac1c$ the operator $A+zB$ is closed on $\cD(A)$ and
\beq\left\{z\in\cc\ |\ |z|<c^{-1}\right\}\ni z\mapsto A+zB\eeq
is a holomorphic family of closed operators. 
  \eet

  \proof
  We easily check that the norms $\sqrt{\|f\|^2+\|Af\|^2}$ and
  $\sqrt{\|f\|^2+\|(A+zB)f\|^2}$ are equivalent for
  $|z|<\frac1c$.
 Let $\cH_0$ be the closure of $\cD(A)$ in $\cH$. The
  restriction of $A$ to $\cH_0$ is densely  defined, so that we can define $A^*$. The operator
$(A^*A+\one)^{-\frac12}$ is unitary from $\cH_0$ to $\cD(A)$. Clearly,
it is bounded in the sense of $\cH_0$. Now
\beq\cc\ni z\mapsto (A+zB) (A^*A+\one)^{-\frac12}\eeq
is obviously a polynomial of degree 1 with values in bounded operators
(hence obviously a holomorphic family). \qed

Let us also quote the following fact proven by Bruk \cite{Bruk}, see also \cite{DW}:
\begin{proposition}\label{adjo} If  $z\mapsto A(z)$ is a holomorphic family of
  closed operators, then so is  $z\mapsto A(\bar z)^*$.
  \end{proposition}

\renewcommand{\abstractname}{Acknowledgements}

\begin{abstract}
  Jan Dereziński is grateful to F.~Gesztesy for drawing
    attention to the references \cite{AlexeevaAnanieva,
      AnanievaBudika, AnanievaBudika2} and for useful comments.  His
    work was supported by National Science Center (Poland) under the
    grant UMO-2019/35/B/ST1/01651.  The authors thank the
    referees for their comments.
\end{abstract}

\end{document}